\documentstyle[version2,aps]{revtex}
\begin{document}
%\draft

\begin{title}
Orbital Correlations in Doped Manganites
\end{title}

\author{
J. P. Hill$^1$, 
C.S. Nelson$^1$, 
M. v. Zimmermann$^1$, 
Y.-J. Kim$^1$,
Doon Gibbs$^1$, D. Casa$^2$, B. Keimer$^3$, Y. Murakami$^4$, C.
Venkataraman$^5$, T. Gog$^5$, Y. Tomioka$^6$, Y. Tokura$^{6,7}$ V.
Kiryukhin$^8$, T.Y. Koo$^8$ and S.-W. Cheong$^{8,9}$ }

\begin{instit}
$^1$ Department of Physics, Brookhaven National Laboratory
Upton, NY 11973-5000

$^2$ Department of Physics, Princeton University
Princeton, NJ 08544

$^3$ Max-Planck-Institut f\"ur Festk\"orperforschung
70569 Stuttgart, Germany

$^4$ Photon Factory, Institute of Materials Structure Science High
Energy Accelerator Research Organization Tsukuba 305-0801, Japan

$^5$ CMC-CAT, Advanced Photon Source, Argonne National Laboratory
Argonne, IL 60439

$^6$ Joint Research Center for Atom Technology (JRCAT) Tsukuba
305-0033, Japan

$^7$ Department of Applied Physics, University of Tokyo, Tokyo
113-0033, Japan

$^8$ Department of Physics and Astronomy, Rutgers University, Piscataway,
NJ 08854

$^9$ Bell Laboratories, Lucent Technologies, Murray Hill, NJ 07974.
\end{instit}

\begin{abstract}

We review our recent x-ray scattering studies of charge and orbital order
in doped manganites, with specific emphasis on the role of orbital
correlations in Pr$_{1-x}$Ca$_x$MnO$_3$. For x=0.25, we find an
orbital structure indistinguishable from the undoped structure and
long range orbital order at low temperatures. For dopings $0.3 \leq x
\leq 0.5$, we find scattering consistent with a charge and orbitally
ordered CE-type structure. While in each case the charge order peaks
are resolution limited, the orbital order exhibits only short range
correlations. We report the doping dependence of the correlation
length and discuss the connection between the orbital correlations and
the finite magnetic correlation length observed on the Mn$^{3+}$
sublattice with neutron scattering techniques. The physical origin of
these domains, which appear to be isotropic, remains unclear. We find
that weak orbital correlations persist well above the phase
transition, with a correlation length of 1-2 lattice constants
at high temperatures.
Significantly, we observe similar correlations at high temperatures in
La$_{0.7}$Ca$_{0.3}$MnO$_3$, which does not have an orbitally ordered
ground state, and we conclude that such correlations are robust to
variations in the relative strength of the electron-phonon coupling.
\end{abstract}

%\pacs{PACS numbers:?}

%]
%\narrowtext
%\twocolumn

%\newpage

\section{Introduction}

The strongly correlated transition metal oxides are characterized by a
wide diversity of ground states, ranging from
antiferromagnetic to ferromagnetic and from insulating to
superconducting. Further, in many cases transitions between these
disparate ground states can be driven by apparently small changes in
some parameter, such as the chemical doping
or temperature. The origin of this dramatic sensitivity 
is believed to lie
in the fact that no single degree of freedom dominates the response,
but rather a number of degrees of freedom may be active. These
can  include the 
spin, charge, lattice and orbital degrees of freedom. 
The ground state is then determined by the interplay between the competing
interests of the relevant degrees of freedom.
However, despite this qualitative understanding, a
complete description of the electronic behavior in transition
metal oxides has proved elusive. Elucidating this behavior remains
one of the central goals in condensed matter physics today.

The perovskite manganites provide an especially illuminating example
of this interplay among the various interactions, since in these
materials all the degrees of freedom are active and the balance 
between them may be conveniently altered \cite{Kaplan-book}. As a
result, much work has been done to understand their magnetic ground
states and lattice distortions, dating back to the early work in the
1950's \cite{Jonkers1,Wollan55}.  However, less is known about
the roles of charge and orbital order in these materials.  The classic
work of Goodenough \cite{Goodenough55}  has nevertheless served as a
guide to their ordered arrangements, as supplemented, for example, by
detailed measurements of the crystal structure and of the temperature
dependence of the lattice constants (see references
\cite{Jirak85,Radaelli97}, for example).

This situation has changed during the last two years following the
detection of orbital and charge order by resonant x-ray scattering
techniques
\cite{Murakami214,Murakami113,Ishihara-PRL-1998,Fabrizio98,Elfimov99,Endoh99,Fulde,Wochner_unpub,Zimmermann99,Zimmermanncond-mat,Paolosini99,Nakao_unpub,Keimer2000,Wakabayashi_unpub,Tanaka99,Hirota2000,Nakao-PRL-2000,Noguchi-YVO,Garcia,Wakabayashi214,McMorrowUPd3,NakaoCeB6,Nakamura99,Hatton99,Benfatto99,Ishihara98}
Specifically,
it has been found that the sensitivity of x-ray scattering to these
structures can be significantly enhanced by tuning the incident x-ray
energy to the transition metal K-absorption edge. Thus, it appears
possible to characterize the orbital and charge ordering on a
microscopic scale, and to study their response to changes of
temperature or to an applied magnetic field. Insofar as we are aware,
resonant x-ray scattering studies of these materials have now been
extended to include La$_{0.5}$Sr$_{1.5}$MnO$_4$ \cite{Murakami214},
LaMnO$_3$ \cite{Murakami113}, La$_{1-x}$Sr$_{x}$MnO$_3$
\cite{Endoh99,Wochner_unpub}, Pr$_{1-x}$Ca$_x$MnO$_3$
\cite{Zimmermann99,Zimmermanncond-mat}, V$_2$O$_3$ \cite{Paolosini99},
YTiO$_3$ \cite{Nakao_unpub}, LaTiO$_3$\cite{Keimer2000},
LaSr$_2$Mn$_2$O$_7$ \cite{Wakabayashi_unpub},
DyB$_2$C$_2$\cite{Tanaka99,Hirota2000},
NaV$_2$O$_5$\cite{Nakao-PRL-2000}, LaVO$_3$ \cite{Noguchi-YVO},
Fe$_3$O$_4$ \cite{Garcia},
La$_{1-x}$Sr$_{1+x}$MnO$_4$ \cite{Wakabayashi214}, 
UPd$_3$ \cite{McMorrowUPd3}, CeB$_6$ \cite{NakaoCeB6}
and
Nd$_{0.5}$Sr$_{0.5}$MnO$_3$\cite{Nakamura99,Hatton99}, and this list
continues to grow.

In this paper, we review our recent x-ray scattering studies of
Pr$_{1-x}$Ca$_x$MnO$_3$ (PCMO) with x=0.25, 0.3, 0.4 and 0.5. Further,
the x=0.3 results are compared with those of La$_{0.7}$Ca$_{0.3}$MnO$_3$
(LCMO system). For PCMO, detailed studies
have been made of the temperature dependence of the orbital and
charge order scattering.
For x=0.3,0.4,0.5, below a doping-dependent ordering
temperature T$_{CO}$, the diffraction pattern is found to be
consistent with the CE-type charge and orbitally ordered structure.
Surprisingly, our
studies reveal that long-range orbital order is never
established in these samples, although long-range
charge order is observed in each case. In contrast, for x=0.25 we
observe only long-range orbital order, consistent with the undoped
structure and with no indication of any charge ordering. Portions of
this work have been published elsewhere
\cite{Zimmermann99,Zimmermanncond-mat,Nelsoncond-mat}

\section{Experimental}

The PCMO crystals used in the present experiments were grown by floating
zone techniques at JRCAT. (0,1,0) surfaces were cut from cylinders of
radius ~3 mm, and polished with fine emery paper and diamond paste.
The mosaic widths of the samples as characterized at the (0,2,0) bulk
Bragg reflections (in orthorhombic $Pbnm$ notation) were typically $\sim$
0.2$^{\circ}$, (FWHM). These values varied by small amounts as the
beam was moved across the surface of each crystals, reflecting its
mosaic distribution. The growth techniques and basic transport
properties of these crystals have been described in detail elsewhere
\cite{Okimoto98,Tomioka96,Tokura}. The LCMO sample was grown by
floating zone techniques at Bell Laboratories and had a similar
mosaic. It was fully twinned with a (110)/(002) surface normal. For
convenience,  we adopt the (110) indexing of the surface-normal 
throughout
this paper \cite{Nelsoncond-mat}.

The x-ray scattering experiments were carried out at the National
Synchrotron Light Source on beamline X22C, and at the Advanced Photon
Source, beamline 9IDB, CMC-CAT. X22C is equipped with a bent, toroidal
focusing mirror and a Ge(111) double crystal monochromator arranged
in a vertical scattering geometry. The optics for 9IDB were comprised
of a double crystal Si(111) monochromator and a flat harmonic
rejection mirror. A  Ge(111) analyzer crystal was employed in the low
temperature
experiments, and a Gr(002) analyzer for the high temperature work.  
All data were taken with the incident x-ray energy
in the vicinity of the Mn K-edge resonance at E=6.555 keV.

\section{Phase behavior of the Pr$_{1-x}$Ca$_{x}$MnO$_3$ system}

At room temperature, the crystal structure of Pr$_{1-x}$Ca$_x$MnO$_3$
is orthorhombic ($Pbmn$).  Characteristic of the perovskite manganites,
each Mn atom lies at the center of the octahedron defined by the
oxygen atoms at the corners.  Single layers of Pr atoms lie between
the layers of octahedra.
A schematic phase diagram for
Pr$_{1-x}$Ca$_x$MnO$_3$ versus Ca concentration and temperature
\cite{Jirak85,Tomioka96} is shown in Figure 1. For small x $(0.15 \leq
x \leq 0.3)$ and at low temperatures, Pr$_{1-x}$Ca$_x$MnO$_3$ is a
ferromagnetic insulator, and is believed to exhibit an 
in plane (a-b) orbitally
ordered ground state analogous to that observed in LaMnO$_3$. The
electronic configuration of the Mn$^{3+}$ (d$^4$) ions is ($t^3_{2g}$,
$e_g^1$) with the $t_{2g}$ electrons localized at the Mn sites. The
$e_g$ electrons are hybridized with the oxygen 2p orbitals, and are
believed to participate in a cooperative Jahn-Teller distortion of the
MnO$_6$ octahedra \cite{Jirak85}. This leads to a
($3x^2-r^2$)-($3y^2-r^2$)-type of orbital order of the $e_g$ electrons
in the ab plane with the oxygens displaced along the direction of
extension of the $e_g$ orbitals. A schematic illustration of this
orbitally ordered state for x=0.25 is shown in Figure 1. The excess
Mn$^{4+}$ ions in this material are believed to be disordered, though
recently, other proposals have been put forward
\cite{Mizokawa2000,Mizokawacondmat00,Kilian1999,Dagotto}. 
To date, however, we have found no evidence of such ordering.
The orbital period is
twice that of the fundamental Mn spacing, so that orbital scattering
appears at structurally forbidden reflections. In orthorhombic
notation, for which the fundamental Bragg peaks occur at (0,2k,0), the
orbital scattering then occurs at (0,k,0).

For Ca concentrations 0.3 $\leq$ x  $\leq$0.7, Pr$_{1-x}$Ca$_x$MnO$_3$
becomes an antiferromagnetic insulator at low temperatures, and
exhibits colossal magnetoresistance in applied magnetic fields, with
the metal-insulator transition occurring between 5 and 8
T\cite{Tomioka96}.  The  insulating phase is accompanied by charge
ordering among the Mn$^{3+}$ and Mn$^{4+}$ ions, and orbital ordering
of the $e_g$ electrons on the Mn$^{3+}$ sites. The large conductivity
results from a delocalization of these $e_g$ electrons and the destruction
of the charge and orbital order. The fraction of Mn ions in the
Mn$^{4+}$ state is determined largely by the concentration of Ca ions.
Thus, by varying the Ca concentration, it is possible to move from a 
ground state with orbital order and no charge order to one in which
both charge and orbital order are observed.
The proposed ground
state\cite{Jirak85} for the $0.3
\leq x \leq 0.7$ concentrations is shown in Figure 1. Note that the same
structure was proposed for  this entire region \cite{Jirak85}.
Clearly, for x $\neq$ 0.5, this picture cannot be strictly correct.
Jirak {\em et al.} proposed that the extra electrons present for
x$\leq$0.5 could be accommodated in such a structure by a partial
occupancy of the $3z^2- r^2$ orbitals of the nominal Mn$^{4+}$ sites.
Other possibilities include small Mn$^{3+}$ rich regions, higher order
structures, or small regions of orbital disorder. As discussed below,
our data reveal that, in fact, the orbital order is not long-ranged in
these compounds, although the charge order is. In the orthorhombic
notation, the charge order reflections occur at (0,2k+1,0) and the
orbital order reflections at (0,k+1/2, 0). Note that the orbital
period ($=2b$) in the x=0.4, 0.5 compounds differs from that occurring
in samples with x$<$0.3 ($=b$), as a result of the presence of charge
ordering.

The magnetic structure of these compounds at low doping (0.15
$\leq$ x $\leq$ 0.3) is ferromagnetic with T$_C \approx$140 K.
Compounds with higher doping (0.3 $\leq$ x $\leq$ 0.75) are
CE-type antiferromagnets with T$_N$=170 K for x between 0.4 and
0.5\cite{Jirak85}.

\section{Low Temperature Correlations}

High-resolution longitudinal scans through the Bragg,
charge and orbital ordering peaks of the x = 0.4 and 0.5 samples
are superimposed on each other for comparison in Figure 2a, and 2b.
These data were obtained at low temperatures (10 K) in the ordered
phase using a Ge(111) analyzer. Solid lines indicate the results of
scans through the (0,2,0) Bragg peaks; open circles indicate scans
through (0,2.5,0) orbital peaks; and filled circles give the results
obtained for the (0,3,0) and (0,1,0) reflections of the charge-ordered
peaks of the x = 0.4 and 0.5 samples, respectively. It is clear from
the figure that the Bragg and charge order peaks have similar widths,
approximately corresponding to the momentum-transfer resolution at
each Q.  This implies that the correlation lengths of the structure
and of the charge order are each at least 2000 $\AA$ 
for both the $x=0.4$
and $x=0.5$ samples. The small differences in width between the
structural and charge order peaks probably reflect the Q-dependence of
the resolution function. In contrast, the orbital ordering peaks in
both samples are significantly broader than the resolution, implying
much smaller orbital domain sizes. We find similar behavior
in x=0.3 samples.

In order to extract longitudinal 
correlation lengths for the orbital order peaks,
we fit these data to a Lorentzian squared lineshape, convolved with a
Lorentzian squared resolution function - the latter being determined
by fits to the (020) structural Bragg peak. This lineshape was chosen
based simply on the quality of the fit,  there is no theoretical
justification for it. 
However, without
an analytical form for the correlation function, we are forced to 
choose a definition for the correlation length, $\xi$.
We take the simplest choice,
$\xi= 1/\Delta k$, where $\Delta k$ is the HWHM of
the Lorentzian-squared. This definition is somewhat generic - given a
particular form for the correlation function, and thus an appropriate
lineshape, the actual correlation length of that model may differ
slightly from those quoted here. However, it is unlikely to
significantly change the results and in the absence of such a
description, we believe that this definition provides a reasonable
empirical characterization of the orbital domain state.

At low temperatures, we find $\xi(x=0.3,$ sample I, the ``disk'' sample)= 60 $\pm 10 \AA
$, $\xi(x=0.3,$ sample II, the ``tombstone'' sample)= 170 $\pm 20 \AA $, $\xi(x=0.4)= 320 \pm
10
\AA $ and $\xi(x=0.5)= 160 \pm 10 \AA $. 
(Note, the two x=0.3 samples
had similar mosaics and the same
reflections were studied. The origin of the differences
in the orbital correlations lengths is not clear). It is important to emphasize
that the associated {\em charge} order of the CE charge and orbitally
ordered state does exhibit significantly longer range 
correlations ($\geq$ 2000$\AA$)
in each case. Thus these results indicate that for these samples an
orbital glass-like state exists  on a well ordered lattice of charge
order. In regard to the concentration dependence of these results, in
an earlier publication \cite{Zimmermann99} we had noted that
$\xi(x=0.5)
\leq \xi(x=0.4)$, despite the fact that the x=0.4 sample did not have 
``enough'' 
Mn$^{4+}$ sites to form the ideal structure. We then speculated
that the shorter correlation length in the x=0.5 structure resulted
from the fact that this sample was closer to tetragonality and that
therefore domain walls which switched $a$ and $b$ axes were more
likely. However, the new $x=0.3$ results do not follow the trend of
longer orbital correlation lengths for more tetragonal samples. This
suggests some other mechanism is controlling the correlation length.
We return to this point below.

The discovery of an orbital domain state sheds light on recent neutron
diffraction studies of Pr$_{0.5}$Ca$_{0.5}$MnO$_3$
\cite{Kajimoto1998,Jirak2}, and powdered
La$_{0.5}$Ca$_{0.5}$MnO$_3$\cite{Radaelli97}. In the PCMO  ($x=0.5$)
it
was shown that the magnetic correlation length was finite. In
La$_{0.5}$Ca$_{0.5}$MnO$_3$, which also exhibits the CE-type magnetic
structure with orbital and charge order\cite{Radaelli97}, separate
magnetic correlation lengths were extracted for the Mn$^{3+}$ and
Mn$^{4+}$ magnetic sublattices, with the remarkable result that they
were quite different: $\xi_{3+}^{mag} = 250-450  \AA$ and
$\xi_{4+}^{mag} \geq 2000 \AA$, respectively.  Those authors proposed
antiphase domain walls composed of ``mis-oriented'' $e_g$ orbitals to
explain the magnetic disorder of the Mn$^{3+}$ sublattice.  Randomly
spaced, domain walls of this type would 
break the orbital coherence but preserve the
charge order coherence (fig. 3). Further, as realized by Radaelli et
al.,  \cite{Radaelli97}, these domain walls affect the magnetic
correlations on the Mn$^{3+}$ and Mn$^{4+}$ sublattice differently.
This is illustrated in fig. 3, in which the signs at each site denote
the spin direction at that site. 
The ideal CE-type charge and orbitally ordered antiferromagnetic
structure is shown in fig. 3a, 
together with various possible domain walls (fig. 3b - 3e). 
Figure 3b corresponds to the domain wall
proposed by Radaelli {\em et al.} \cite{Radaelli97}
with an orbital phase shift running perpendicular to 
the orbital propagation vector.
Inspection of fig. 3b reveals
that the magnetic coherence on
the Mn$^{3+}$ sublattice is broken by this domain wall, while the
Mn$^{4+}$ sublattice remains unaffected. Such orbital domain walls
would therefore explain the observed magnetic neutron diffraction
data. Our observation of orbital correlation lengths of similar values
to the Mn$^{3+}$ magnetic correlations strongly suggests that Radaelli
$et~al.$ were correct in their speculation, and that we have observed
these antiphase domains directly in PCMO. Note that these domains are
believed to be static, and do not correspond  to the (dynamic) orbital
fluctuations inferred from magnetic neutron diffraction investigations
of the ferromagnetic spin fluctuations in Pr$_{1-x}$Ca$_x$MnO$_3$,
which disappear below T$_N$ \cite{Kajimoto1998}.

If all of the domain walls were of the type  shown in fig 3b, however,
then the orbital coherency would only be broken along one direction,
and very anisotropic domains would be observed. To look for such
anisotropy, 
we measured the
correlation lengths in the other two directions, i.e. perpendicular to
{\bf b$^{*}$}, the orbital propagation direction. These measurements
were carried out on the x=0.3 sample I, the ``disk'' sample, with a Ge(111) analyzer and
are shown in fig. 4. We find that the orbital domains, in this sample,
are approximately isotropic, with slightly reduced correlation lengths
in the two transverse directions, of $\xi_H= 36 \AA$ and $\xi_L=46
\AA$ along the H and L directions respectively (Note that these two correlation
lengths correspond to widths significantly broader than the 
resolutions 
in these two directions, which are determined by the sample mosaic and
the out-of-plane collimation, respectively. Thus no correction for resolution effects
was made). These results indicate the 
presence of domain walls, other than those of fig. 3b, that run in
other directions and which disrupt only the orbital correlations. 

Before discussing this further, we point out that it
appears that this orbital glass-like state is common in 
manganites with the CE-type 
charge and orbital structure - we have observed it in
all PCMO samples studied, the La$_{0.5}$Ca$_{0.5}$MnO$_3$ neutron
diffraction data indicate that is also present in this compound
\cite{Radaelli97}, and very recent x-ray work on 
(La$_y$Pr$_{1-y}$)$_{1-x}$Ca$_{x}$MnO$_3$ revealed the presence of a
similar orbital glass/charge ordered state in that system as well
\cite{Kiryukhincond-mat}. 
%Finally, in La$_{0.67}$Ca$_{0.33}$MnO$_3$
%an incommensurate state is observed. The incommensurability is
%believed to be the result of an ordered array of domain walls
%(discommensurations) breaking the coherence of an otherwise
%commensurate CE-type structure. The spacing of these domain walls can
%be obtained from the incommensurability and is 200$\AA$ ({\bf ???
%check this. Also is the charge order comm or IC, ie do these domain
%%walls screw up the CO? If so they are of a different type to those
%discussed here. But still it'll be worth mentioning.})

In each of these cases, orbital domain walls spaced by a few hundred
angstroms are observed in an ordered array of Mn$^{3+}$ and Mn
$^{4+}$ sites. It is interesting to ask what are the
underlying energetics that might determine this length scale. There
are a number of different contributions to the cohesive energy of the
CE state shown in fig. 3a (in the third dimension an identical charge
and orbitally ordered layer is stacked on top of the one shown in fig.
3a, with all the spins reversed) \cite{Goodenough55,Yunoki-PRL,Brink-PRL}. 
The first of
these are the magnetic bonds, both ferromagnetic and
antiferromagnetic, resulting from the presence or absence of occupied
orbitals between the various Mn sites, and mediated by the
oxygens \cite{Goodenough55}. In the antiferromagnetic state shown in
fig. 3a, all these are satisfied, that is none of the bonds are
frustrated. A second contribution is the Coulomb
energy of the Mn charges. This is minimized by the in-plane charge ordered
lattice shown. Third, the CE structure contains a series of
ferromagnetic zig-zag stripes oppositely aligned and running through
the $ab$ plane (fig. 3a). The $e_g$ electrons are free to hop along these
zig-zags as in the case of a double-exchange-like ferromagnet. This
allows them to increase their kinetic energy in this
antiferromagnetic insulating state. Finally there is the electronic energy
gain associated with the local Jahn-Teller distortion around each
Mn$^{3+}$ which removes the degeneracy of the $3z^2-r^2$ and $x^2-y^2$
orbitals (so-called ``non-cooperative phonons''\cite{Yunoki-PRL}) 
and the gain in
energy associated with the orbital coherency which results from the 
fact that neighboring Mn sites share oxygens (``cooperative phonons''
\cite{Yunoki-PRL}). These last two come 
at the expense of lattice energy resulting
from the associated oxygen motion.

In figures 3b)-e), we compare possible domain walls in the $ab$ plane
that break the orbital order coherence, but preserve the charge order
coherence. The first thing to note is that in each case, all the 
magnetic bonds are satisfied, that is there is no magnetic frustration
introduced by these domain walls. Secondly, in each case, the magnetic
coherence of the Mn$^{4+}$ lattice is left undisturbed, while that on the
Mn$^{3+}$ is broken. In the first two, figs. 3b and c, the orbital
coherence is broken along the $b$ and $a$ directions respectively, but
in each case the propagation vector is left unchanged - along $b$.
In the second two cases, figs. 3d and 3e, the orbital correlations
are again broken along the $b$ and $a$ directions, respectively. However,
in these cases, the propagation vector is rotated by 90$^{\circ}$ across
the domain wall. 

In light of the energy considerations discussed above, 
we see that the domain walls of
the type shown in fig. 3b cost very little energy.  They do not break
any magnetic bonds, and preserve the ferromagnetic zig-zag stripes,
simply adding an extra straight section to one of the ``zigs'' (though there
is an energy price, relative to the ground state, associated with 
straight lines of 1D ferromagnetism \cite{Yunoki-PRL}). The
charge order coherence is not disturbed, so there is no Coulomb
price to pay relative to the ground state and there are the same
number of local distortions as there are in the ideal structure. The
principal cost then, relative to the ideal structure is that due to the
fact that along the domain wall, the oxygen motions are not cooperative.
Similar arguments can be made about the domain walls shown in fig 3c. For
cases d) and e), the argument is slightly different, since
in each case, in addition
to the energy cost of the domain wall itself, there is also some energy
cost associated with the part of the sample that has the ``wrong''
orientation of the propagation vector, i.e. along a local a-axis.
For PCMO, x=0.5, however, the difference between a and b is very small
 \cite{Jirak2} and thus the energy cost might also be expected to be 
small.

If the orbital twins shown in fig. 3d and 3e coincide with structural
twins - that is the $a$ and $b$ crystallographic axes are also interchanged
across these walls - then we can rule out these domain walls as 
limiting the orbital correlations. This is because the structural coherence,
as measured at the (020) reflection would also be limited by such domain
walls, and yet, as  we have seen,
it exhibits significantly longer correlations than those
of the orbital peaks (figure 2). However, if such orbital
twins exist in a $single$ crystallographic domain, then they would
have the desired properties of limiting the orbital coherence, but leaving
the charge order and structural coherence unchanged.

The experimental signature of such orbital
twins would be a (0.5,0,0)-type orbital
peak in a single crystallographic domain sample. In a 
crystallographically twinned sample,
it would be difficult to distinguish such peaks from a (0,0.5,0)-type 
peak associated with an $a$-axis crystallographic twin. 
However, in PCMO $x=0.3$ (the ``tombstone'' sample),
we were able to clearly resolve the (200) and (020) reflections, along
the nominal (0k0) direction,
from $a$ and $b$ twins. In this sample, in addition to the 
(0,0.5,0)-type peaks associated
with the b-axis twin, we also observed (0.5,0,0)-type peaks associated
with the a-axis twin. Such peaks are consistent with the presence of
orbital
domains of the type shown in fig. 3d and 3e in a single crystallographic
domain and are inconsistent with the ideal structure or with 
orbital domain walls of 
the type shown in fig. 3b and 3c, which don't affect the propagation 
vector. 

The question remains, however, what sets the length scale for the
domain wall separation? That is, what is the energy gain associated 
with the insertion of domain walls? One possibility is that there is an
impurity potential arising from the presence of Ca ions in the structure.
Local fluctuations in the concentration could pin the orbital
order on a particular sublattice and introduce domain walls of the type
shown in fig. 3. While such a scenario is  hard to  rule out, it does 
not naturally explain the few hundred angstrom domain size observed. 
In addition, the Pr and Ca ions are extremely similar in size -
Pr$^{3+}$ is 1.126$\AA$ and Ca$^{2+}$ is 1.12$\AA$, so any local strains
from dopants are expected to be minimal \cite{Frontera-PRB-2000}.

A second possibility arises from the fact that with 
the formation of the CE charge and
orbitally ordered state, comes a significant  change in the lattice 
constants, which dramatically increases the orthorhombicity. Thus, if
each crystallographic domain were to be a single orbital domain it would
have to accommodate a significant strain associated with the increase 
in $a$ and $b$ lattice constants. However, if it were to break up into
a number of orbital domains with propagation vectors in each of the
three directions, (i.e. domain walls of type in fig 3d and fig 3e, 
together with a third in the out-of-plane direction), then 
the strain would be minimized in that volume. It may be that 
balancing the build-up of strain energy associated with the 
orbital order  with the energy cost of domain walls of this type
is what is determining the few hundred angstrom domain size 
\cite{Millis_private}. Testing this hypothesis will require detailed
energy calculations for the CE charge and orbitally ordered state,
the cost of the domain wall and the build up of the long range strain.
It is hoped that these speculations will prompt such calculations.

\section{High Temperature Correlations}

We now turn our attention  to a discussion of the high temperature
correlations, i.e those observed well above the phase transition. We
will compare the correlations observed in two manganites that have
very different ground states, specifically the PCMO x=0.3 (``tombstone'') 
sample and
La$_{0.7}$Ca$_{0.3}$MnO$_3$ (LCMO). These are isostructural compounds,
both exhibiting P$bnm$ symmetry with the same formal valence on the Mn
site. The only difference between the two is the size of the
rare-earth ion: La is about 3$\%$ bigger than Pr. This has the effect
of decreasing the distortion of the Mn-O-Mn bond angles, bringing it
closer to the ideal 180$^{\circ}$ in the LCMO compound, thus
increasing the electronic bandwidth and the elastic modulus. Both
these effects decrease the relative strength of the electron-phonon
coupling \cite{millis1,roder,egami} and as a result, LCMO does not
charge and orbitally order, but rather undergoes a transition from
an insulating state into a
metallic ferromagnetic state below $T_{\rho}$=250 K. By comparing the
correlations observed well above the respective phase transitions of
these two x=0.3 manganites, we aim to shed light on the role of the
relative strength of the
electron-phonon coupling.

We begin with the La$_{0.7}$Ca$_{0.3}$MnO$_{3}$ sample.  At
temperatures above the metal-insulator transition temperature, broad
peaks with ordering wavevectors of (0.5 0 0) and (0 0.5 0) and peak
intensities of $\sim$20 counts/s (on beamline X22C at the NSLS) were
observed.  Note that twinning of the sample and the width of the
diffuse peaks make it impossible to determine whether or not there is
a unique ordering wavevector, however, these wavevectors are consistent
with the peaks arising from orbital correlations of the CE-type
structure. 
Representative scans at two
temperatures, 260 K and 220 K, are shown in Figure 5.
As the sample was cooled through the transition temperature
into the ferromagnetic metallic phase, the peaks abruptly decreased in
intensity (see inset to Figure 5).

Scans were performed along both H and K directions in reciprocal space
and the resulting data fit to Lorentzian squared lineshapes, as
outlined above. The correlations lengths obtained in this manner were
found to be temperature independent and approximately isotropic, with
a magnitude of 1-2 lattice constants in both directions. The
temperature dependence of the fitted values from the H scans 
are shown in figure 6, in which the temperature is
recorded as a reduced temperature $t=(T-T_{\rho})/T_{\rho}$.

The temperature dependence of the orbital correlations observed in
PCMO x=0.3 are also shown in figure 6, again as a function of
reduced temperature, where in this case $T_{\rho}$ is replaced by
$T_{CO}=200 K$. On warming from below the transition, the HWHM for the
PCMO sample is observed to increase rapidly above the transition.
However, this broadening does not continue indefinitely, rather the
HWHM is observed to saturate around 40K above T$_{CO}$. Significantly,
the value that it saturates at is the same as that observed in the
LCMO, and again corresponds to 1-2 lattice constants.

Thus in these two dissimilar manganites we find very similar high
temperature correlations. Specifically, they have the same wavevector,
size and temperature dependence (at least far from the phase
transition where the intensity in each case increases as the samples
are cooled). We suggest that because the same correlations are observed
in such different manganite systems, they must be robust to variations
in the relative strength of the electron-phonon coupling and are
therefore likely to be common to a large class of manganites. This in
turn suggests that there may be something fundamental about these
correlations that makes them particularly stable. We further speculate
that these correlations are in fact small regions of orbital order, as
shown in figure 7 \cite{Nelsoncond-mat}. A similar picture has been
proposed previously \cite{Kim2000}.

There are a number of pieces of evidence that support this conclusion.
First, the wavevector of the scattering is consistent with such a
structure (and inconsistent with other possible descriptions,
including so-called ``orbital polarons'
\cite{Kilian1999,Mizokawacondmat00}). Second, these correlations are
observed to evolve continuously into the ordered CE structure in the
PCMO x=0.3 case. Third, the magnetic interactions within this small
piece are all ferromagnetic. Thus, this picture of the orbital
correlations is consistent with the  neutron scattering observation of
ferromagnetic fluctuations in PCMO above T$_{CO}$ \cite{Kajimoto1998}.
The fact that they are ferromagnetic also serves to make them
particularly robust since this lessens the energy price
associated with localizing the electrons.

There are a number of possible descriptions that could be applied to
these correlations. One could term them bipolarons to emphasize the
fact that they represent a pairing of two polaronic distortions,
around each Mn$^{4+}$ site.  An alternative label would be
``ferromagnetic zig-zags'' to emphasize the connection with the ordered
CE-type structure, or finally one could refer to them as
ferromagnetic clusters - to draw parallels with the phase separation
picture of the metal-insulator transition in the manganites. It will
require more experiments to determine which of these labels, if any,
is the more appropriate. For example, investigating the doping
dependence, applying a magnetic field or going to
yet higher temperatures would shed light on the stability of these
correlations against various perturbations and thus help elucidate their
nature.

\section {Summary}

We have used x-ray scattering techniques to study the orbital
correlations in a number of doped manganites. In the PCMO
series, for x=0.25, we find a ground state with long range orbital
order and no evidence for charge order. For $0.3 \leq x \leq 0.5$,
for which the low temperature ground state is the charge and
orbitally ordered CE-type antiferromagnetic structure, the orbital
correlations do not develop long range order, but rather form a domain
state. These correlations are approximately isotropic, and domain
sizes are on the order of a few hundred Angstroms. In contrast, the
charge ordering in this structure does exhibit long range order.
The evidence suggests that orbital domain walls are a common feature of
the CE-type manganites, however, there is as yet no theoretical
understanding of their origin.

In addition, we have  studied the correlations observed at very high
temperatures, well above the phase transition. Here, we observe short
range orbital correlations of 1-2 lattice constants in extent.
Further, similar correlations are observed in two very different
manganites, PCMO (x=0.3) and LCMO (x=0.3), which have
antiferromagnetic, charge and orbitally ordered insulating, and
ferromagnetic metallic ground states, respectively. We suggest that the
presence of similar correlations in such dissimilar manganites
demonstrates the robustness of these correlations to changes in the
relative strength of the electron-phonon coupling and we present an
intuitively appealing description of these correlations.

\section{ Acknowledgements}

We acknowledge helpful conversations with S. Ishihara, D.J. Khomskii,
S. Maekawa, A.J. Millis, and G. A. Sawatzky. The work at Brookhaven, both in the
Physics Department and at the NSLS, was supported by the U.S. Department
of Energy, Division of Materials Science, under Contract No.
DE-AC02-98CH10886, and at Princeton University by the N.S.F. under Grant
No. DMR-9701991. Support from the Ministry of Education, Science and
Culture, Japan, by the New Energy and Industrial Technology Development
Organization (NEDO), and by the Core Research for Evolution Science and
Technology (CREST) is also acknowledged. Work at the CMC beamlines is
supported, in part, by the Office of Basic Energy Sciences of the U.S.
Department of Energy and by the National Science Foundation, Division of
Materials Research.  Use of the Advanced Photon Source was supported by
the Office of Basic Energy Sciences of the U.S. Department of Energy
under Contract No. W-31-109-Eng-38.

%\bibliography{PCMO}

%\bibliographystyle{Prsty}

\figure{ Composition-temperature phase diagram of
Pr$_{1-x}$Ca$_x$MnO$_3$ in zero magnetic field (following
reference \cite{Jirak85}). The full lines indicate the charge/orbital
transition temperature ($T_{OO/CO}$), antiferromagnetic transitions ($T_{N}$)
are marked with dashed lines and ferromagnetic transitions ($T_C$) with
dotted lines. The two insets represent schematics of the in-plane orbital
and charge ordered
structures. Only Mn ions are represented. The elongated shapes represent
3$z^2-r^2$ orbitals on Mn$^{3+}$ sites. The circles represent Mn$^{4+}$ 
sites.}

\figure{ Upper: Longitudinal scans of the Bragg (0,2,0), the
charge (0,1,0), and the orbital (0,2.5,0) reflections of the x=0.4
sample at T=8 K. The secondary peaks to the right of the charge order
and Bragg reflections arise from structural twins.
Lower: The same for the x=0.5 sample. Data have
been normalized to the same peak intensity to facilitate
comparison.}

\figure{ Schematic of the CE-type charge and orbitally ordered antiferromagnetic
state and possible domain walls. 
Signs indicate spin components at each Mn site. a)
ideal structure. b)-e) various domain walls which preserve the charge order
but not the orbital coherence. Domain walls in b) and c) preserve the 
orbital propagation vector (0,0.5,0) and those in d) and e) rotate it by
90$^{\circ}$.
Note that in each case, the magnetic coherence of the 
Mn 3+ sublattice has been
broken, relative to a), but not that of the Mn 4+
sublattice. The solid lines outline ferromagnetic stripes in these
structures. In each case, the b crystallographic axis runs vertically in
the figure, the a axis horizontally. }

\figure{
Scans through the (0,2.5,0) orbital reflection in
Pr$_{0.7}$Ca$_{0.3}$MnO$_3$, sample I,  along each of the three 
orthorhombic axes in reciprocal
space. The correlation lengths quoted in each figure were extracted from
fits to the data, as discussed in the text.}

\figure{Reciprocal space scans along H, with Lorentzian-squared
fits (line), in La$_{0.7}$Ca$_{0.3}$MnO$_{3}$.  Data were measured at temperatures
of 260 K (open) and 220 K (closed).  The peaks at (1.67 2 0) and (1.5 2.13 0)
arise from powder lines and were excluded from the fits.  Inset shows
temperature dependence of scattering intensity at (1.5 2 0) (open) and (1.3 2 0)
(closed).  Note that spurious points at temperatures between 270 and 275 K
coincided with a beam dump.}

\figure {The fitted HWHM values of the (1.5 2 0) scattering as a function of
reduced temperature ($\textstyle t\equiv\frac{T-T_{\rho/co}}{T_{\rho/co}}$),
in La$_{0.7}$Ca$_{0.3}$MnO$_{3}$ (open) and Pr$_{0.7}$Ca$_{0.3}$MnO$_{3}$ (closed).}

\figure {Schematic diagram of the proposed structure of the observed high-temperature
correlations, in the $a-b$ plane.  Open circles represent Mn$^{4+}$
ions; elongated figure-eights represent the occupied e$_{g}$
(3d$_{z^{2}-r^{2}}$) orbital of Mn$^{3+}$ ions; closed circles
represent Mn ions that, on average, have the formal valence and no net
orbital order; and arrows indicate the in-plane component of the
magnetic moment.}

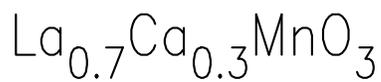
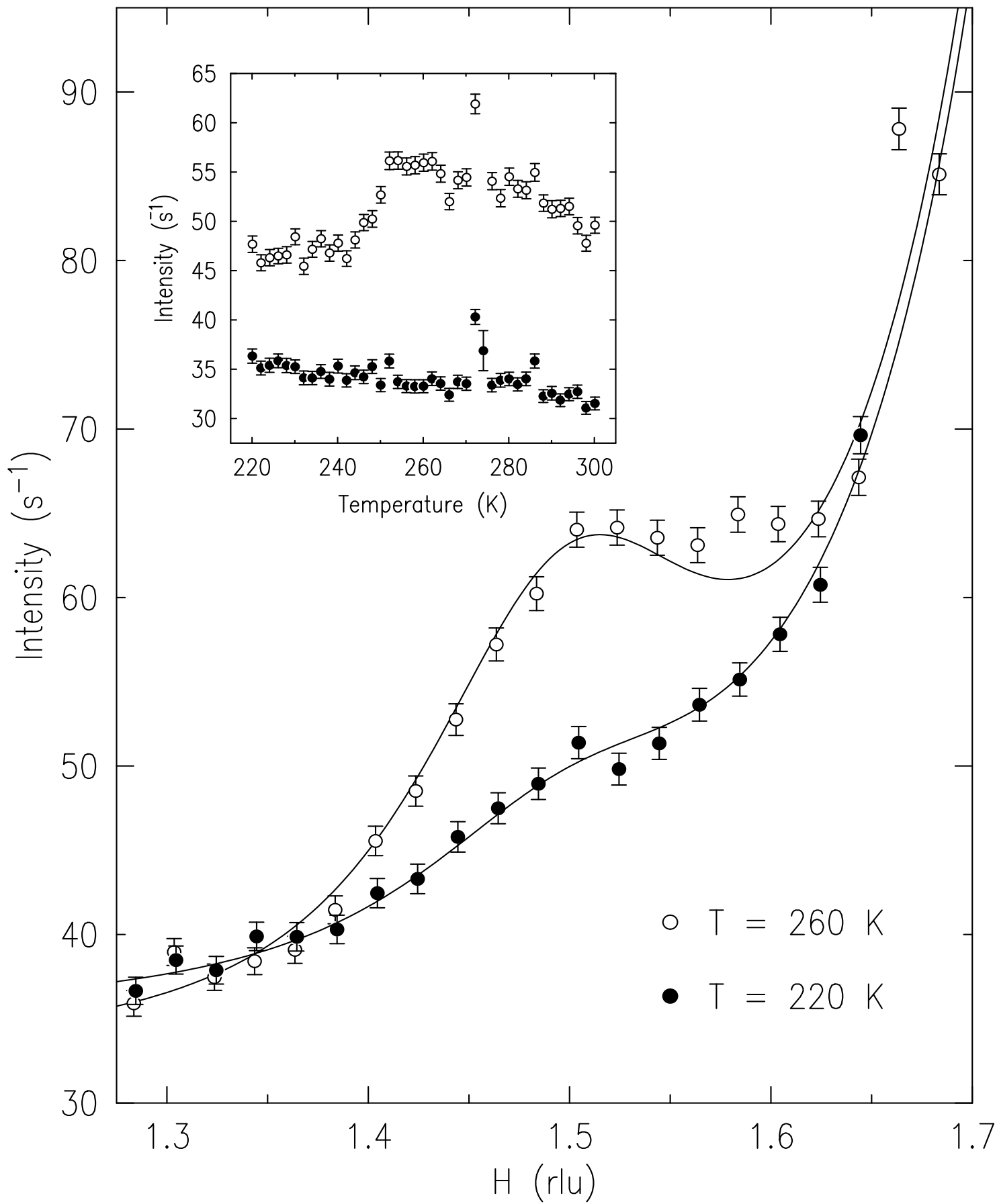
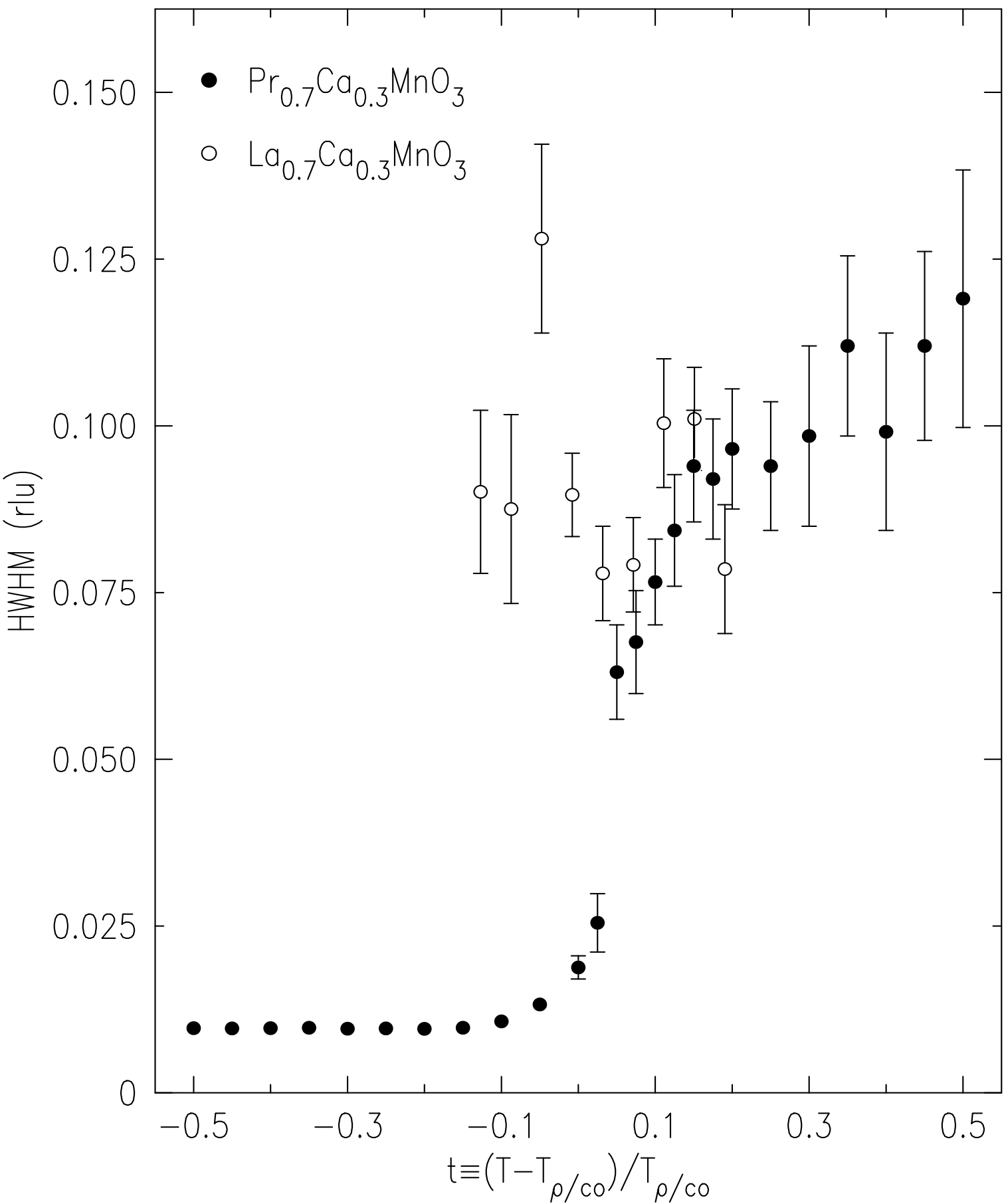
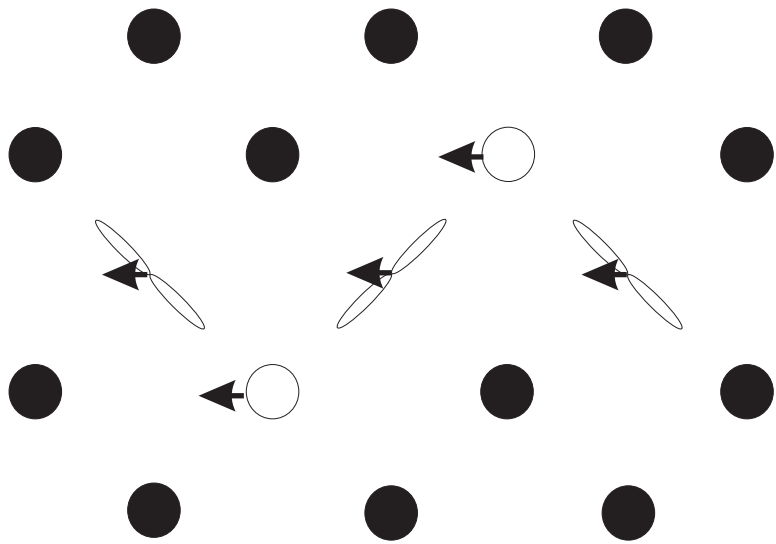
\end{document}